\documentclass[aps,prl,twocolumn,nofootinbib,showpacs,floatfix,superscriptaddress,preprintnumbers]{revtex4}
\usepackage{amsmath}
\usepackage{amssymb}
\usepackage{epsfig}
\usepackage{graphicx}
\usepackage{pstricks}
\usepackage{pst-coil}
\DeclareMathAlphabet{\mathsc}{OT1}{cmr}{m}{sc}
\newcommand {\ignore}[1]{}

\def\10{$SO(10)$}
\def\21{SU(2) $\otimes$ U(1) }

\def\422{$SU(4) \otimes SU(2) \otimes SU(2)$}
\def\321{SU(3) $\otimes$ SU(2) $\otimes$ U(1)}
\def\gsim{\raise0.3ex\hbox{$\;>$\kern-0.75em\raise-1.1ex\hbox{$\sim\;$}}}
\def\lsim{\raise0.3ex\hbox{$\;<$\kern-0.75em\raise-1.1ex\hbox{$\sim\;$}}}

\def\lsim{\raise0.3ex\hbox{$\;<$\kern-0.75em\raise-1.1ex\hbox{$\sim\;$}}}
\def\gsim{\raise0.3ex\hbox{$\;>$\kern-0.75em\raise-1.1ex\hbox{$\sim\;$}}}
\def\vev#1{\left\langle #1\right\rangle}
\def \znbb {0\nu\beta\beta}

\newcommand{\AddrAHEP}{%
  AHEP Group, Institut de F\'{\i}sica Corpuscular --
  C.S.I.C./Universitat de Val{\`e}ncia \\
  Edificio Institutos de Paterna, Apt 22085, E--46071 Valencia, Spain}

 \newcommand{\ba}{\begin{array}}
\newcommand{\ea}{\end{array}}
\relax

\def\321{$SU(3)\times SU(2)\times U(1)$}

\begin{document}

\preprint{IFIC/10-19}

\renewcommand{\Huge}{\Large}
\renewcommand{\LARGE}{\Large}
\renewcommand{\Large}{\large}
\def \znbb {$0\nu\beta\beta$ }
\def \nbb {$\beta\beta_{0\nu}$ }
\title{Discrete dark matter}  
\author{M.~Hirsch}\email{mahirsch@ific.uv.es}  
\author{S.~Morisi}\email{ morisi@ific.uv.es}
\author{E.~Peinado}\email{epeinado@ific.uv.es}
\author{J.~W.~F.~Valle} \email{valle@ific.uv.es}
\affiliation{\AddrAHEP}

\date{\today}

\begin{abstract}

  We propose a new motivation for the stability of dark matter
  (DM). We suggest that the same non-abelian discrete flavor symmetry
  which accounts for the observed pattern of neutrino oscillations,
  spontaneously breaks to a $Z_2$ subgroup which renders DM stable. 
  The simplest scheme leads to a scalar doublet DM potentially detectable
  in nuclear recoil experiments, inverse neutrino mass hierarchy,
  hence a neutrinoless double beta decay rate accessible to upcoming
  searches, while $\theta_{13}=0$ gives no CP violation in neutrino
  oscillations.

\end{abstract}


\maketitle

\textbf{Introduction}
The existence of dark matter (DM) plays a central role in the modeling
of structure formation and galaxy evolution, affecting also the cosmic
microwave background. Despite the strong evidence in favor of DM, its
detailed nature remains rather elusive.
Viable particle physics candidates for dark matter must be
electrically neutral, and provide the correct relic
abundance. Therefore they must be stable over cosmological time
scales.
A simple way to justify the stability of the DM is by {\it assuming}
some parity symmetry $Z_2$, which might arise from the spontaneous
breaking of an abelian U(1) gauge symmetry
\cite{Frigerio:2009wf,Kadastik:2009dj,Batell:2010bp}~\footnote{In
  supersymmetry a viable DM particle is the neutralino, whose
  stability stems from the imposition of the so-called R-parity. }, or
from a non-abelian discrete symmetry, as might be the case in some
string models~\cite{dine:2007}.

Non abelian discrete symmetries are motivated by neutrino oscillation
data \cite{babu:2002dz,altarelli:2005yp}.  Here we propose that the
same symmetry explaining neutrino mixing angles is also responsible
for the dark matter stability.
In our simplest type-I seesaw~\cite{valle:2006vb} realization the
flavor symmetry $A_4$ spontaneously breaks to $Z_2$ providing a stable
DM candidate.  We extend the scalar sector of the standard model by
adding three Higgs doublets transforming as a triplet of $A_4$ we show
that there is a consistent pattern of vacuum expectation values (vevs)
for which only one of the three extra Higgs doublets takes a vev,
while the other two give rise to the dark matter candidate. The model
accounts for the observed pattern of mixing
angles~\cite{schwetz:2008er} indicated by current neutrino oscillation
data, predicting $\theta_{13}=0$ and inverted spectrum of neutrino
masses. It will therefore be tested in upcoming double beta and
neutrino oscillation searches~\cite{nunokawa:2007qh}, while the dark
matter
has  potentially detectable rates within reach of nuclear recoil experiments.\\[-.3cm]

\textbf{Model}
We assign matter fields to irreducible representations of $A_4$, the
group of even permutations of four objects, isomorphic to the symmetry
group of the tetrahedron. All elements are generated from two elements
$S$ and $T$ with $S^2=T^3=(ST)^3=\mathcal{I}$.
$A_4$ has four irreducible representations, three singlets
$1,~1^\prime$ and $1^{\prime \prime}$ and one triplet.  In the basis
where $S$ is real diagonal,
\begin{equation}\label{eq:ST}
S=\left(
\begin{array}{ccc}
1&0&0\\
0&-1&0\\
0&0&-1\\
\end{array}
\right)\,;\quad
T=\left(
\begin{array}{ccc}
0&1&0\\
0&0&1\\
1&0&0\\
\end{array}
\right)\,;
\end{equation}
one has the following triplet multiplication rules, 
\begin{equation}\label{pr}
\begin{array}{lll}
(ab)_1&=&a_1b_1+a_2b_2+a_3b_3\,;\\
(ab)_{1'}&=&a_1b_1+\omega a_2b_2+\omega^2a_3b_3\,;\\
(ab)_{1''}&=&a_1b_1+\omega^2 a_2b_2+\omega a_3b_3\,;\\
(ab)_{3_1}&=&(a_2b_3,a_3b_1,a_1b_2)\,;\\
(ab)_{3_2}&=&(a_3b_2,a_1b_3,a_2b_1)\,,
\end{array}
\end{equation}
where $\omega^3=1$, $a=(a_1,a_2,a_3)$ and $b=(b_1,b_2,b_3)$.  We
assign the standard model Higgs doublet $H$, to the singlet $1$, and
we assume three additional Higgs doublets transforming as an $A_4$
triplet, namely $\eta=(\eta_1,\eta_2,\eta_3)\sim 3$.  We have four
right-handed neutrinos, three transforming as an $A_4$ triplet
$N_T=(N_1,N_2,N_3)$, and one singlet $N_4$. The lepton and Higgs
assignments of our model is in table\,\ref{tab1}.
\begin{table}[h!]
\begin{center}
\begin{tabular}{|c|c|c|c|c|c|c|c|c||c|c|}
\hline
&$\,L_e\,$&$\,L_{\mu}\,$&$\,L_{\tau}\,$&$\,\,l_{e}^c\,\,$&$\,\,l_{{\mu}}^c\,\,$&$\,\,l_{{\tau}}^c\,\,$&$N_{T}\,$&$\,N_4\,$&$\,H\,$&$\,\eta\,$\\
\hline
$SU(2)$&2&2&2&1&1&1&1&1&2&2\\
\hline
$A_4$ &$1$ &$1^\prime$&$1^{\prime \prime}$&$1$&$1^{\prime \prime}$&$1^\prime$&$3$ &$1$ &$1$&$3$\\
\hline
\end{tabular}\caption{Summary of  relevant  model quantum numbers}\label{tab1}
\end{center}
\end{table}
The resulting Yukawa Lagrangian is
\begin{eqnarray}\label{lag}
\mathcal{L}&=&y_e L_el_{_e}^c H+y_\mu L_\mu l_{_\mu}^c H+y_\tau L_\tau l_{_\tau}^c H+\nonumber\\
&&+y_1^\nu L_e(N_T\eta)_{1}+y_2^\nu L_\mu(N_T\eta)_{1''}+y_3^\nu L_\tau(N_T\eta)_{1'}+\nonumber\\
&&+y_4^\nu L_e N_4 H+ M_1 N_TN_T+M_2 N_4N_4+
\mbox{h.c.}\nonumber
\end{eqnarray}
This way $H$ is responsible for quark and charged lepton masses, the
latter automatically diagonal.
Note that we do not discuss the quark sector, assumed to be blind to
$A_4$, namely all left and right-handed up and down-type quarks
transform trivially under $A_4$, their mass and mixing hierarchies
might arise from an extra family symmetry, for example,
Frogatt-Nielsen-like~\cite{Froggatt:1978nt}.  Neutrino masses arise
from $H$ and $\eta$, see below.
The relevant terms of the scalar potential are of the form
\begin{equation}
\begin{array}{lll}
V&=&\mu_\eta^2\eta^\dagger\eta
+\mu_H^2 H^\dagger H+\lambda_1 [H^\dagger H]^2+\lambda_2 [\eta^\dagger\eta]_1^2+\\
&+&\lambda_3 [\eta^\dagger\eta]_{1^\prime}[\eta^\dagger\eta]_{1^{\prime\prime}}
+\lambda_4 [\eta^\dagger\eta^\dagger]_{1^\prime}[\eta\eta]_{1^{\prime\prime}}+\lambda_4' [\eta^\dagger\eta^\dagger]_{1}[\eta\eta]_{1}\\
&+&\lambda_5\sum_i[\eta^\dagger\eta]_{3_{i}}[\eta^\dagger\eta]_{3_{i}}
+\lambda_5'([\eta^\dagger\eta]_{3_{1}}[\eta^\dagger\eta]_{3_{2}}+h.c.)+\\
&+&\lambda_6(\sum_{i,j}[\eta^\dagger\eta^\dagger]_{3_{i}}[\eta\eta]_{3_{j}}+h.c)+\lambda_7 [\eta^\dagger\eta]_1  H^\dagger H +\\ 
&+&\lambda_7' [\eta^\dagger H]  H^\dagger \eta +\lambda_8 \left([\eta^\dagger\eta^\dagger]_1  H H+h.c\right)+
\\&+&\lambda_9 \left([\eta^\dagger\eta]_{3_{1}}  \eta^\dagger H+h.c\right)+
\lambda_9' \left([\eta^\dagger\eta]_{3_{2}}  \eta^\dagger H+h.c\right)+\\
&+&\lambda_{10} \left([\eta^\dagger\eta^\dagger]_{3_{1}}  \eta H+h.c\right)+\lambda_{10}' \left([\eta^\dagger\eta^\dagger]_{3_{2}}  \eta H+h.c\right)+\\
&+&\lambda_{11} [\eta^\dagger\eta^\dagger]_1[\eta\eta]_1
\label{potential}
\end{array}
\end{equation}
where $i,j=1,2$, and $[...]_{3_{i}}$ means the product of two triplets
contracted into a triplet of $A_4$, see eq.\,(\ref{pr}), $[...]_{1}$
means the product of two triplets contracted into a singlet of $A_4$
and so on.  Note that $[\eta^\dagger \eta]_{1,1',1''}\equiv
[\eta\eta^\dagger]_{1,1',1''}$, $[\eta \eta]_{3_1}\equiv
[\eta\eta]_{3_2}$ and so on.\\

%
We have studied the minimization of the potential $V$ solving the equations
$\partial V/\partial v_i=0$ where $v_i$ are the vevs of the fields $H,
\eta_1,\eta_2$ and $\eta_3$.  For simplicity we assume real vevs.
We have checked that for suitable parameter choices of the potential
$V$, an allowed  local minimum is
\begin{equation}
  \vev{ H^0}=v_h\ne 0,~~~~ \vev{ \eta^0_1}=v_\eta \ne 0~~~~
\vev{ \eta^0_{2,3}}=0\,,
\end{equation}
with the eigenvalues of the Hessian $\partial^2 V/\partial v_i \partial
v_j$ all positive.

Note that the alignment $\vev{ \eta} \sim (1,0,0)$ breaks
spontaneously $A_4$ to $Z_2$ since $(1,0,0)$ is invariant under the
$S$ generator in eq.~(\ref{eq:ST}).
The $Z_2$ is defined as
\begin{equation}\label{residualZ2}
\begin{array}{lcrlcrlcr}
N_2 &\to& -N_2\,,\quad& h_2 &\to& -h_2\,, \quad&A_2 &\to& -A_2\,, \\   
N_3 &\to& -N_3\,,\quad& h_3 &\to& -h_3\,,\quad &A_3&\to& -A_3\,.  
\end{array}
\end{equation}
This residual symmetry is responsible for the stability of our DM
candidate and the stability of the minimum.  Note that the potential
cannot break spontaneously $A_4$ into $Z_3$ because in this case the
alignment $\vev{ \eta} \sim (1,1,1)$ is not a minimum unless a fine
tuning in the parameters $\lambda_9+\lambda_{10}=0$ is assumed.  This
atractive feature reminds of the inert doublet
model~\cite{Barbieri:2006dq}, with the difference that here it follows
naturally from the underlying flavor symmetry which accounts for
neutrino oscillations.

We have four Higgs doublets\footnote{ Lepton flavor violating
  processes are suppressed by the large right-handed neutrino scale.}
giving three physical charged scalar bosons, plus four neutral
scalars, and three pseudoscalars. After electroweak symmetry breaking
we can write
\begin{equation}\begin{array}{cc}
 H=\left(
\begin{array}{c}
0\\
v_h+h
\end{array}
\right),
&\eta_1=\left(
\begin{array}{c}
\eta_1^+\\
v_\eta+h_1+iA_1
\end{array}
\right),\\ \\
\eta_2=\left(
\begin{array}{c}
\eta_2^+\\
h_2+iA_2
\end{array}
\right),&\eta_3=\left(
\begin{array}{c}
\eta^+_3\\
h_3+iA_3
\end{array}
\right).
\end{array}
\end{equation}
There are 3 physical charged scalar bosons, 4 CP even and 3 CP odd
  neutral scalars.   The mass of the neutral scalar fields is block
diagonal with the standard model Higgs $h$ mixed with $h_1$, but not
with the scalar fields with zero vev's $h_{2,3}$.\\[-.3cm]

\textbf{Dark matter}
The lightest combination of the stable scalar fields $h_2$, $h_3$ 
plays the role of our dark matter particle, which we will denote
generically by $\eta_{DM}$.
We list below all interactions of $\eta_{DM}$:
\begin{enumerate}
\item Yukawa interactions
\begin{equation}
\begin{array}{l}
\eta_{_{DM}}\, \overline{\nu}_i   N_{2,3}\,,
\end{array}
\end{equation}
where $i=e,\,\mu,\,\tau$.
\item Higgs-Vector boson couplings
\begin{equation}\label{eq:gint}
\begin{array}{l}
\eta_{_{DM}}\eta_{_{DM}} ZZ\,,\quad \eta_{_{DM}}\eta_{_{DM}} WW\,,\\
\eta_{_{DM}}\eta_{2,3}^{\pm} W^\pm Z\,,\quad \eta_{_{DM}}\eta_{2,3}^{\pm} W^\pm \,,\\
\eta_{_{DM}}A_{2,3} Z \,.
\end{array}
\end{equation}
\item Scalar interactions from the Higgs potential:
\begin{equation}\label{eq:Pint}
\begin{array}{l}
\eta_{_{DM}}\, A_1 A_2 h\,,\quad \eta_{_{DM}}\, A_1 A_3 h_1\,,\\
\eta_{_{DM}}\, A_1 A_2 h_1\,,\quad \eta_{_{DM}}\, A_1 A_3 h\,,\\ 
\eta_{_{DM}}\, A_2 A_3 h_3\,,\quad\eta_{_{DM}}\, h_1 h_3 h\,\\ 
\eta_{_{DM}}\eta_{_{DM}} hh\,,\quad\eta_{_{DM}}\eta_{_{DM}} h_1h_1\,.
\end{array}
\end{equation}
\end{enumerate}
After electroweak symmetry breaking, the vevs $v_h$ and $v_\eta$ are
generated, so that additional terms are obtained from those in
Eq.~(\ref{eq:Pint}) by replacing $ h\to v_h$ and $h_1\to v_\eta$. The
flavor symmetry $A_4$ is broken down to the residual $Z_2$ symmetry in
Eq.~(\ref{residualZ2}), implying the stability of our dark matter
candidate. As we will see, despite the many mass and coupling
  parameters appearing in the potential, eq. (\ref{potential}), for
  $M_\eta\gg M_z$, only two determine the relic dark matter abundance
  and its direct
  detection rates.\\[-.2cm]

\textbf{Relic Density}
Assuming that our DM candidate arises as thermal relic in the early
universe, one of the most important requirements one must check is its
relic abundance. For definiteness we require that $\eta_{DM}$ makes up
all the observed DM.
For $M_\eta\gg M_z$ the most important annihilation and coanihilation
processes are those with vector bosons, though for large $\lambda
\gsim g_2$, where 
$16\lambda^2=(\lambda_7+\lambda_7'+2\lambda_8)^2+(2\lambda_2-\lambda_3-2\lambda_4+\lambda_4'+2\lambda_5+\lambda_5'+\lambda_6)^2$, annihilation into
Higgs bosons plays an important role, see Eq.~(\ref{relic2}).  The DM
abundance can be approximated as~\cite{Cirelli:2005uq}
\begin{equation}
\label{relic1}
\frac{n_{\rm DM}(T)}{s(T)} \approx \sqrt{\frac{180}{\pi\ g_*}}\frac{1}{M_{\rm Pl}\ T_f \langle \sigma_A v\rangle},
\end{equation}
where $\frac{M_\eta}{T_f} \approx 26$ and $g_*=106.75+n$ is the
number of SM degrees of freedom plus $1 \leq n \leq 12$ degrees of
freedom arising from the extra scalars, and $M_{Pl}=1.22\times 10^{19}
GeV$ is the Planck scale.  The cross section for
$\eta_{DM}\eta_{DM}\to VV$ where $V$ are vector bosons in the limit of massless final states, is given
by~\cite{Cirelli:2005uq}
\begin{equation}\label{relic2}
  \langle \sigma_A v\rangle \simeq\frac{3 g_2^4  +  g_Y^4 + 6 g_2^2 g_Y^2 Y^2 +4 \lambda^2}{256 \pi\ M_\eta^2},
\end{equation}
where $Y=1/2$ is the weak hypercharge, $g_2=\sqrt{4
  \pi \alpha/(1-M_W^2/M_Z^2)}$ and $g_Y=\sqrt{4 \pi \alpha }
M_Z/M_W$.
From these equations it follows that, in order to provide the correct
relic abundance $\Omega_{\rm DM} h^2=0.110\pm 0.006$ i.e.\ $n_{\rm
  DM}/s = (0.40\pm0.02)eV/M_\eta$~\cite{komatsu:2008hk}, a correlation
between the mass of the dark matter $M_\eta$ and the quartic coupling
constant $\lambda$ is required.  For simplicity if we take the limit
of small $\lambda$ we obtain a mass for the DM candidate of $
M_\eta\approx 0.51 ~\mbox{TeV}.$ For large $\lambda$ values we have
that the DM mass $M_{\eta}$ scales as $\lambda$.\\[-.2cm]

\begin{figure}
\begin{center} 
\includegraphics[width=4.2cm]{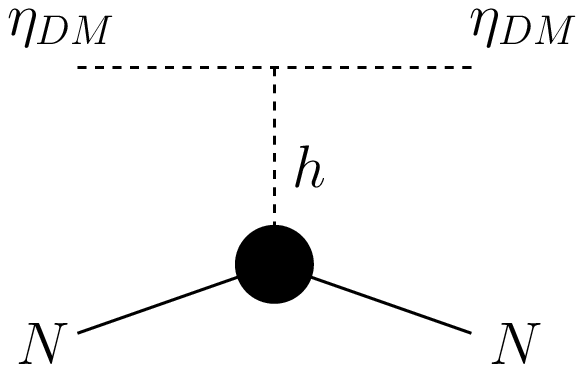}
\includegraphics[width=4.2cm]{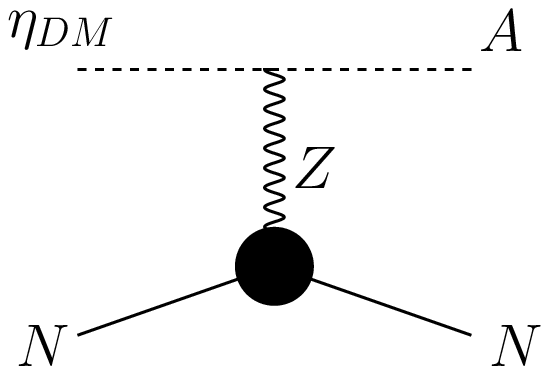}
\caption{Feynman diagrams relevant for direct DM detection. Elastic
  scattering (left) is generically more important than inelastic
  (right).}
\label{fig1}
\end{center}
\end{figure}

\textbf{Direct detection}
The quartic couplings $\eta^\dagger\eta H^\dagger H$ and
$\eta^\dagger\eta^\dagger H H$ give an interaction of the DM candidate
with the nucleon through the interchange of the SM Higgs boson.  Hence
our DM candidate can be detected through the elastic scattering with a
nucleus $\eta_{DM} N\to \eta_{DM} N$ via the exchange of a Higgs, or
through inelastic scattering with a nucleus $\eta_{DM} N \to A N$ with
the exchange of a $Z$ boson, see Fig.~\ref{fig1}, where $A$ is the
lightest pseudoscalar, in general a mixture of $A_2$ and $A_3$.
  
\begin{center} 
\begin{figure}[h!] 
\vglue -1cm
\includegraphics[width=9.5cm]{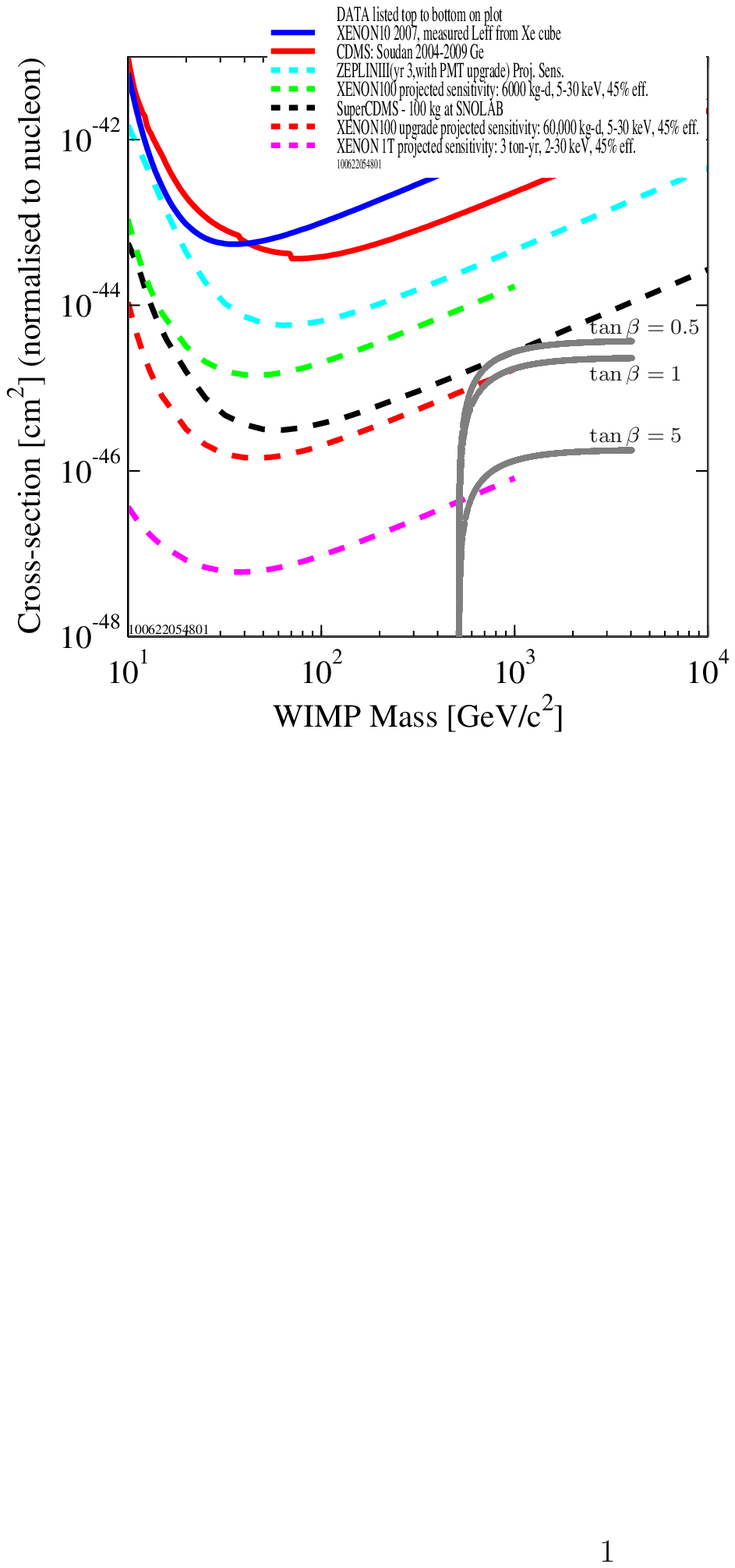}
\vglue -.5cm
\caption{Elastic DM scattering cross section with a nucleon versus DM
  mass.  We compare present \cite{Angle:2007uj,Ahmed:2009zw} and
  future \cite{Aprile:2010um,Aprile:2009yh} sensitivities with our
  model expectations, for $m_H=120$\,GeV and $\tan \beta=0.5,~1,~5$
  (grey solid lines). } \vglue -.8cm
\label{fig2}
\end{figure}
\end{center}

Barring fine-tuned choices of parameters for which the threshold for
inelastic scattering opens up, the detection will be dominated by the
elastic process, whose cross section is given by~\cite{Burgess:2000yq}
\begin{eqnarray}\label{directdet}
&&\sigma_{\rm el}(\hbox{nucleon}) \approx \lambda^2\, \frac{1}{1+(\tan\beta)^{2}}\left( \frac{100~{\rm GeV}}{M_h} \right)^4\times 
\nonumber\\
&&\qquad\qquad\times 
\left( \frac{50~{\rm GeV}}{M_\eta} \right)^2 ~ \Bigl( 5 \times 10^{-42}
~{\rm cm}^2 \Bigr)  \,,\nonumber\\
&&
\label{sisN}
\end{eqnarray}
where $\tan \beta=v_h/v_\eta$.  Note that all uncertainties associated
with the nuclear form factor in Eq.~(\ref{sisN}), have been neglected.
From the requirement of correctly reproducing the relic density,
Eqs. (\ref{relic1}) and (\ref{relic2}), one can find an expression for
$\lambda$ as function of the DM mass, $M_\eta$. Using this relation
and eq. (\ref{directdet}) one can plot the estimated cross section for
the direct detection for each value of $\tan \beta$ and mass of the
Higgs, $M_h$, as illustrated in Fig.~\ref{fig2}~\footnote{Here we
  focus on the region $M_\eta\gg M_z$. The interesting case of light
  DM will be treated elsewhere.  }.  The figure has been generated
using \cite{plot} and compares the experimental sensitivities with our
model expectations, fixing $m_H=120$~GeV and three $\tan \beta$
values. This choice of Higgs mass is motivated by the LEP bounds
$m_H>114$\,GeV, which however is not strictly valid in our model due
to the additional Higgs doublets.
\textbf{Neutrino phenomenology}
Our model has four heavy right-handed neutrinos, and is a special
case, called (3,4), of the general type-I seesaw
mechanism~\cite{schechter:1980gr}.  After electroweak symmetry
breaking, it is characterized by Dirac and Majorana mass terms given
as
\begin{equation}
m_D=\left(
\begin{array}{cccc}
x_1&0&0&y_1\\
x_2&0&0&0\\
x_3&0&0&0\\
\end{array}
\right),
M_R=\mathrm{diag}(M_1,M_1,M_1,M_2)~,
\end{equation}
so that the light neutrinos get Majorana mass by means of the type-I
seesaw relation
$
m_\nu=-m_{D_{3\times 4}}M_{R_{4\times 4}}^{-1}m_{D_{3\times 4}}^T
$
the light-neutrinos mass matrix $M_{\nu}$  being given as
\begin{equation}
\label{mnu}
\left(
\begin{array}{ccc}
\frac{x_{1}^2}{M_{1}}+\frac{y_{1}^2}{M_{2}}&\frac{x_{1}x_{2}}{M_{1}}&\frac{x_{1}x_{3}}{M_{1}}\\
\frac{x_{1}x_{2}}{M_{1}}&\frac{x_{2}^2}{M_{1}}&\frac{x_{2}x_{3}}{M_{1}}\\
\frac{x_{1}x_{3}}{M_{1}}&\frac{x_{2}x_{3}}{M_{1}}&\frac{x_{3}^2}{M_{1}}
\end{array}
\right)
=\left(
\begin{array}{ccc}
y^2& ab&ac\\
ab&b^2&bc\\
ac&bc&c^2
\end{array}
\right).
\end{equation}
It falls within the class of scaling matrices introduced in Ref.\,\cite{Mohapatra:2006xy}.
This form of the light neutrino mass matrix has an inverse
hierarchical neutrino mass spectrum and a zero eigenvalue with $m_3=0$
and corresponding eigenvector $(0,\,-c/b,\,1)^T$ implying a vanishing
reactor mixing angle $ \theta_{13}=0$.  One can see explicitly that
the solar and atmospheric square mass differences and mixing angles
indicated by neutrino oscillation data~\cite{schwetz:2008er} can
indeed be fitted by taking, as an example, the tri-bimaximal (TBM)
\emph{ansatz} \cite{Harrison:2002er}.  When $b=c$ and $y^2=2c^2-ac$
the neutrino mass matrix Eq.~(\ref{mnu}) is $\mu-\tau$ invariant
yielding maximal atmospheric mixing, $\sin^2\theta_{23}=1/2$ and
$M_{\nu,11}+M_{\nu,13}=M_{\nu,22}+M_{\nu,23}$, which gives the TBM
value of the solar angle, $\sin\theta_{12}^2=1/3$, in good agreement
with experimental data within one~$\sigma$. The eigenvalues are $
\{m_1,m_2,m_3 \}=\{2ac+2c^2,2c^2-ac,0 \}\,, $ which can fit the two
mass-squared differences required to account for the observed pattern
of neutrino oscillations.  By relaxing the condition $b=c$ and
$y^2=2c^2-ac$ one generates deviations from the TBM limit, while
keeping $\theta_{13}=0$.  Note the above imples a neutrinoless double
beta decay effective mass parameter in the range 0.03 to 0.05~eV at
3~$\sigma$, within reach of upcoming
experiments~\cite{avignone:2007fu}.\\\vspace{-.2cm}

\textbf{Conclusions}
In summary we have suggested that DM stability follows from the same
non-abelian discrete flavor symmetry which accounts for the observed
pattern of neutrino oscillations. In the realization we have
  given we have an $A_4$ symmetry which spontaneously breaks to a
$Z_2$ parity that stabilizes a scalar doublet dark matter, potentially
detectable in nuclear recoil experiments, as well as accelerators.
 Despite the complexity of the scalar potential, in the heavy
  dark matter limit both the relic dark matter abundance and its
  direct detection cross section depend just on the DM mass and a
  single coupling strength parameter.
  The model is also manifestly unifiable and agrees with electroweak
  searches as well as precision tests, as will be shown elsewhere.
Our simple example gives \znbb rates accessible to upcoming
experiments and no CP violation in neutrino oscillations.\\[-.2cm]

\textbf{Acknowledgments}
We thank Marco Taoso for useful discussions.
This work was supported by the Spanish MICINN under grants
FPA2008-00319/FPA and MULTIDARK Consolider CAD2009-00064, by
Prometeo/2009/091, by the EU grant UNILHC PITN-GA-2009-237920.
S.~M. is supported by a Juan de la Cierva contract.



\end{document}